\documentstyle{aipproc}

\input psfig

\def\xslide#1#2#3#4#5#6#7{\centerline{\psfig
{figure=#1,height=#2,bbllx=#3bp,bblly=#4bp,bburx=#5bp,bbury=#6bp,width=#7,clip=}}}

\title{Analysis of New Results For Scalar-Isoscalar \pp Phase Shifts}

\vspace{-0.2cm}

\author{R. Kami\'nski$^*$, L. Le\'sniak$^*$ and B.\ Loiseau$^{\dagger}$}

\vspace{-0.4cm}

\address{$^*$Department of Theoretical Physics, 
H. Niewodnicza\'nski Institute of Nuclear Physics,\\PL 31-342
Krak\'ow, Poland\\
$^{\dagger}$Division de Physique Th\'eorique,\\
Institut de Physique Nucl\'{e}aire,  
F-91406, Orsay Cedex\\ 
and LPTPE Universit\'e P. \& M. Curie, 4 Place Jussieu,
             F-75252, Paris Cedex 05, France}

\newcommand {\bfq}{{\bf q}}
\newcommand {\bfp}{{\bf p}}

\newcommand {\la}{\lambda}
\newcommand {\eq}{\begin{equation}}
\newcommand {\qe}{\end{equation}}
\newcommand{\ba}{\begin{eqnarray}}
\newcommand{\ea}{\end{eqnarray}}
\newcommand{\pp}{$\pi\pi$ }
\newcommand{\kk}{$K\overline{K}$ }
\newcommand{\fo}{$f_0(980)$ }
\newcommand{\epw}{$f_0(1400)$ }
\newcommand{\epsig}{$f_0(500)$ }

\newcommand{\reactpol}{$\pi^- p_{\uparrow} \rightarrow \pi^+ \pi^- n$ }

\newcommand{\roro}{$\sigma\sigma$ }

\begin{document}

\maketitle

\vspace{-0.8cm}

\begin{abstract}
The scalar -- isoscalar \pp 
phase shifts are analysed using 
a separable potential model of three coupled channels
(\pp, \kk and an effective $2\pi 2\pi$ system).
Model parameters are fitted to two sets of solutions obtained in a recent
analysis of the CERN-Cracow-Munich measurements of the \reactpol
reaction on a polarized target.
A relatively narrow (90 -- 180 MeV) scalar resonance $f_0(1400-1460)$ is found,
together with a wide \epsig ($\Gamma \approx 500$ MeV) and the narrow \fo state.

\end{abstract}

\vspace{-0.5cm}

Recently the CERN-Cracow-Munich data
[1]
for the $\pi^-p_{\uparrow} \to \pi^+\pi^-n$ reaction
on a polarized target were analysed in Ref. 
[2].
Four solutions for
the isoscalar $S$-wave phase shifts from the \pp threshold up to 1600 MeV
were found and two of them, 
called "down-flat" and "up-flat" fully satisfy unitarity constraints.
The "down-flat" solution is in good agreement with the former solution
of Ref. [3] below 1400 MeV. 
In Ref. [4] the scalar meson spectrum was studied in terms of
a relativistic \pp and \kk coupled channel 
model from the $\pi\pi$ threshold up to 1400 MeV. 
Strong four--pion production, observed in different experiments 
[5--9],
provides a compelling argument to take into
account the $4\pi$ channel. 
In this report we extend the isoscalar $S$--wave 2--channel
model of Ref. [4] by adding to its $\pi\pi$ and $K\bar K$
channels an effective third coupled channel, here
called $\sigma\sigma$.

We consider the $S$-wave scattering and transition reactions between three 
coupled channels of meson pairs labelled 1, 2 and 3. Reaction amplitudes 
$T$ satisfy a system of coupled channel Lippmann-Schwinger equations
[4] with a separable form of the interaction:
\vspace{-0.3cm}
\eq <\bfp|V_{\alpha \gamma}|\bfq> = \sum_{j=1}^{n}\la_{\alpha \gamma,\,j}\ 
                             g_{\alpha,\,j}(\bfp)\ g_{\gamma,\,j}(\bfq),                  
                              \ \ \alpha, \gamma = 1, 2, 3,
                              \label{pot}\qe                              
where $\la_{\alpha \gamma,\,j}$ are coupling constants 
and 
$g_{\alpha,\,j}(\bfp)= (4\pi/m_j)^{1/2}/
                             (\bfp^2+\beta_{\alpha,\,j}^2)$
are form factors which depend on the relative centre of mass meson 
momenta $\bfp$ in the final channel or $\bfq$ in the initial channel.  In the
\pp channel ($\alpha,\gamma=1$) we choose a rank-2 separable potential ($n=2$) 
and in the other channels, i.e.  $K \overline{K} \ (\alpha,\gamma= 2)$ 
and $\sigma \sigma \ (\alpha,\gamma=3)$, a rank-1 potential ($n=1$).

The model has 14 parameters: 9 coupling constants
$\la_{\alpha \gamma,\,j}$, 4 range
parameters $\beta_{\alpha,\,j}$ and the $\sigma$
mass $m_3$.
We can solve the system of Lippmann-Schwinger equations
following the formalism developed in Refs. [4,10].
The $S$-matrix elements $S_{\alpha \beta}\ (\alpha, \beta = 1, 2, 3)$ can be 
written in terms of the Jost function of different arguments (see 
[10]).
The model satisfies the unitarity condition $S^+S = 1$.
Some of the $S$-matrix poles in the
complex energy plane can be interpreted as resonances. 
The diagonal matrix elements are parametrized as
$S_{\alpha\alpha} = \eta_{\alpha}e^{2i\delta_{\alpha}}$,
where $\eta_{\alpha}$ and $\delta_{\alpha}$ are the channel $\alpha$ 
inelasticities and phase shifts, respectively. 
We fit the new experimental results 
[2] 
on the
$\pi\pi$ $S$-wave isoscalar phase shifts and 
inelasticities together with  the $K\bar K$ phase shifts from
[11]. 

At the beginning we have considered only 
the $\pi\pi$ and $K\bar K$ 2--channel case with 8 free  parameters. 
Next the starting parameters for 3--channel case were taken from 
the 2--channel fits.
In Fig. 1 two different fits (A and B) to the "down-flat" data for 
$\delta_{\pi\pi},
\eta_{\pi\pi}$ and $\varphi_{\pi K}=
\delta_{\pi\pi}+\delta_{KK}$ are compared to the experiment.
In Fig. 1a we have only shown the fit A since the energy dependence
of the fit B is very close to that of A. 
Fits to the "up-flat" data are shown in 
[10].
We found that in the 2--channel case for the "down-flat" solution it was not
possible to obtain a good fit to the $\eta$ values between 1400 to 1600 MeV.
In the 3--channel model, however, we can get a substantial
decrease of $\eta$ above 1400 MeV (see Fig. 1b). 
In order to achieve this behaviour, couplings between the $\pi\pi$ and
$\sigma\sigma$ or \kk channels should be sizable. 
The main difference between the 2-- and 3--channel fits lies in $\eta$
above 1400 MeV, where the opening of the 
$\sigma\sigma$ channel
leads to a fast decrease of inelasticity parameters.
Let us also note an improvement in $\varphi_{\pi K}$ over the 1000 to 1200 MeV 
range, as it can be seen in Fig. 1c.
The corresponding parameters for these best fits are given in Table
\ref{parameters}. 

Fits of similarly good quality were obtained with very 
different physical pa\-ra\-me\-ters in the \kk and \roro channels.
For example, in the 2--channel fits and 
in fit A the \kk interaction is attractive while 
in the case B it is repulsive (see Table \ref{parameters}). 
Similarly, interchannel couplings are very different in both cases.
In fit A we see rather strong \pp to \roro  and \kk to \roro
couplings, while in the case B the \pp -- \kk coupling, $\Lambda_{12,2}$,
is particularly strong.

We have studied positions of the $S$--matrix poles in the complex energy plane
($E_{pole} = M- i\Gamma/2$).
For the 3--channel model there are 8  different sheets
which correspond to different signs of imaginary
parts of the channel momenta ($Im p_1, Im p_2, Im p_3$).
For example, on the sheet denoted by ($---$) all imaginary parts are negative.
Resonance parameters predicted by the 3--channel model are summarized in
Table 2.
At low energy we find a very broad \epsig resonance
(also called $\sigma$ meson).
The \fo resonance is seen in the vicinity of
the \kk threshold with a width of about 60 to 70 MeV.
Mass of a relatively narrow state \epw varies from about 1400 MeV to 1460 MeV.
For the "down--flat" fits this resonance is narrower
($\Gamma \approx 100$ MeV) on sheet $(--+)$ than on sheet $(---)$.
The parameters of this resonance are close to those of the $f_0(1500)$
resonance - a hypothetical glueball state-found by the Crystal Barrel Group
[5,6] in $p\overline{p}$ annihilation.

\vspace{0.1cm}

This work has been performed in the framework of the IN2P3 -- 
Polish Laboratories Convention (project No 93-71) and partially supported by the 
Polish State Committee for Research (grants No 2P03B 231 08 and 2 P03B 020 12).

\begin{figure}[h!]
\xslide{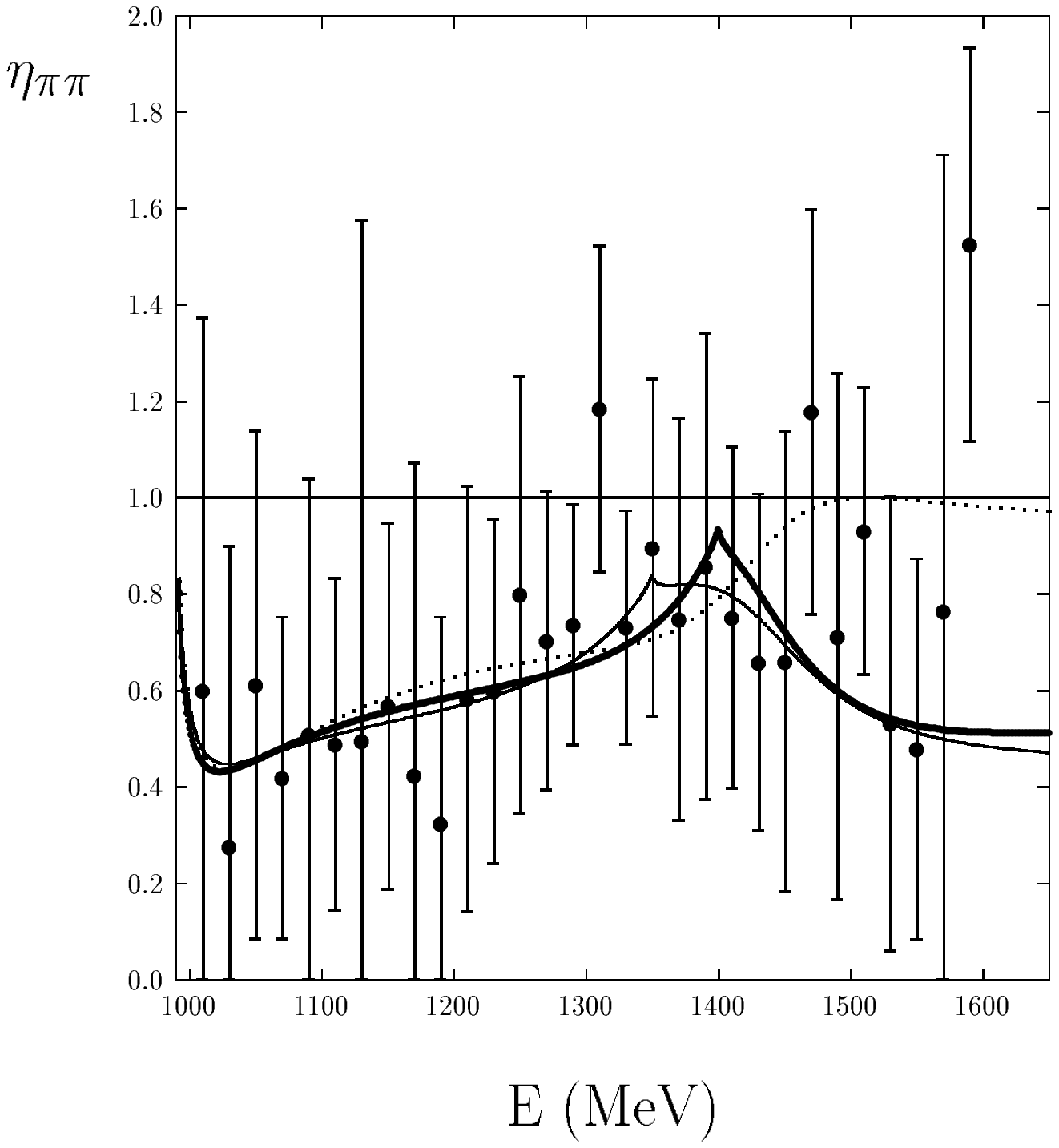}{7cm}{-300}{190}{505}{600}{15cm} 
\vspace{-7cm}
\hspace{-3.5cm}\xslide{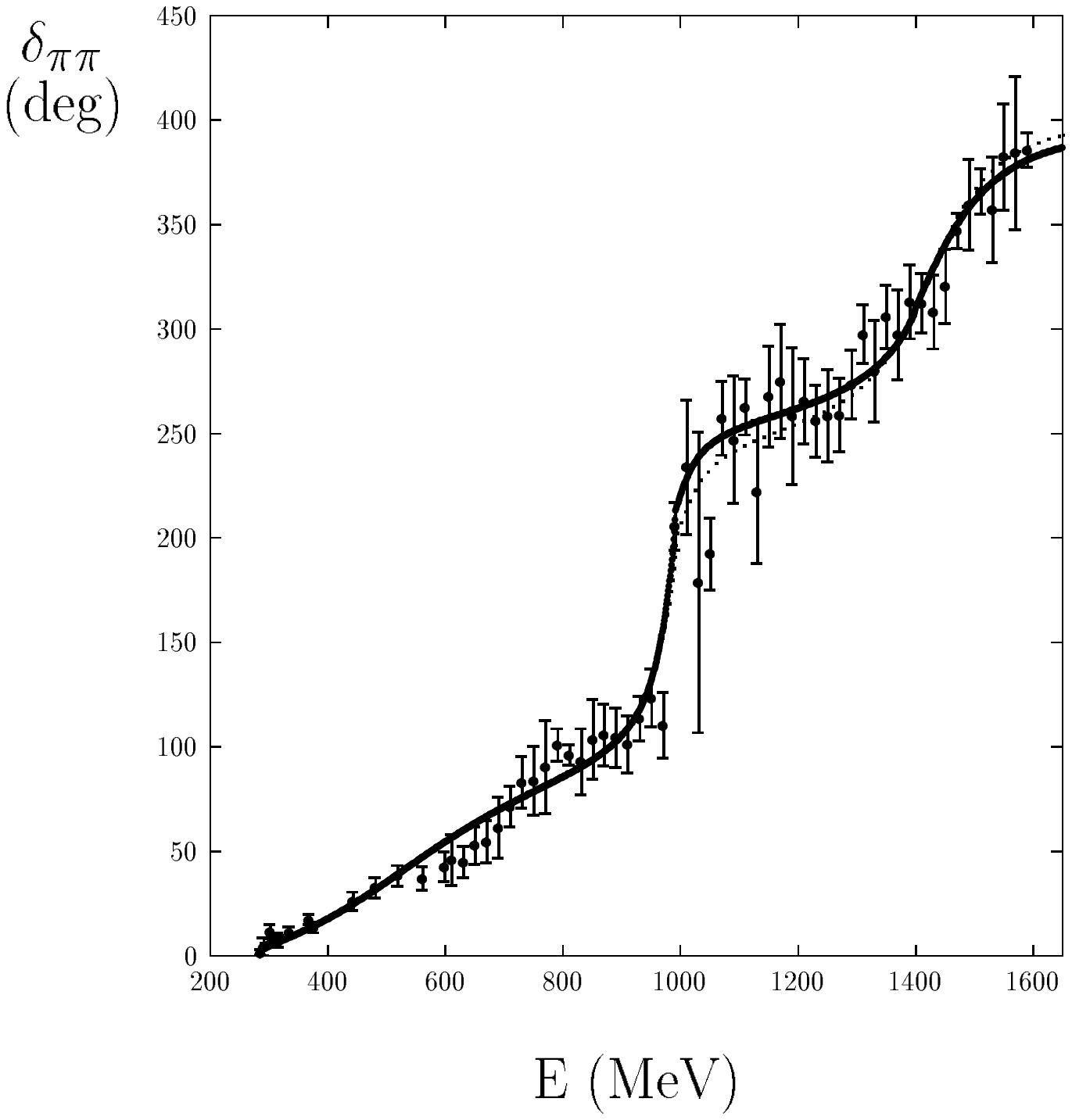}{7cm}{100}{185}{505}{600}{7.7cm} 
\vspace{-0.2cm}
\xslide{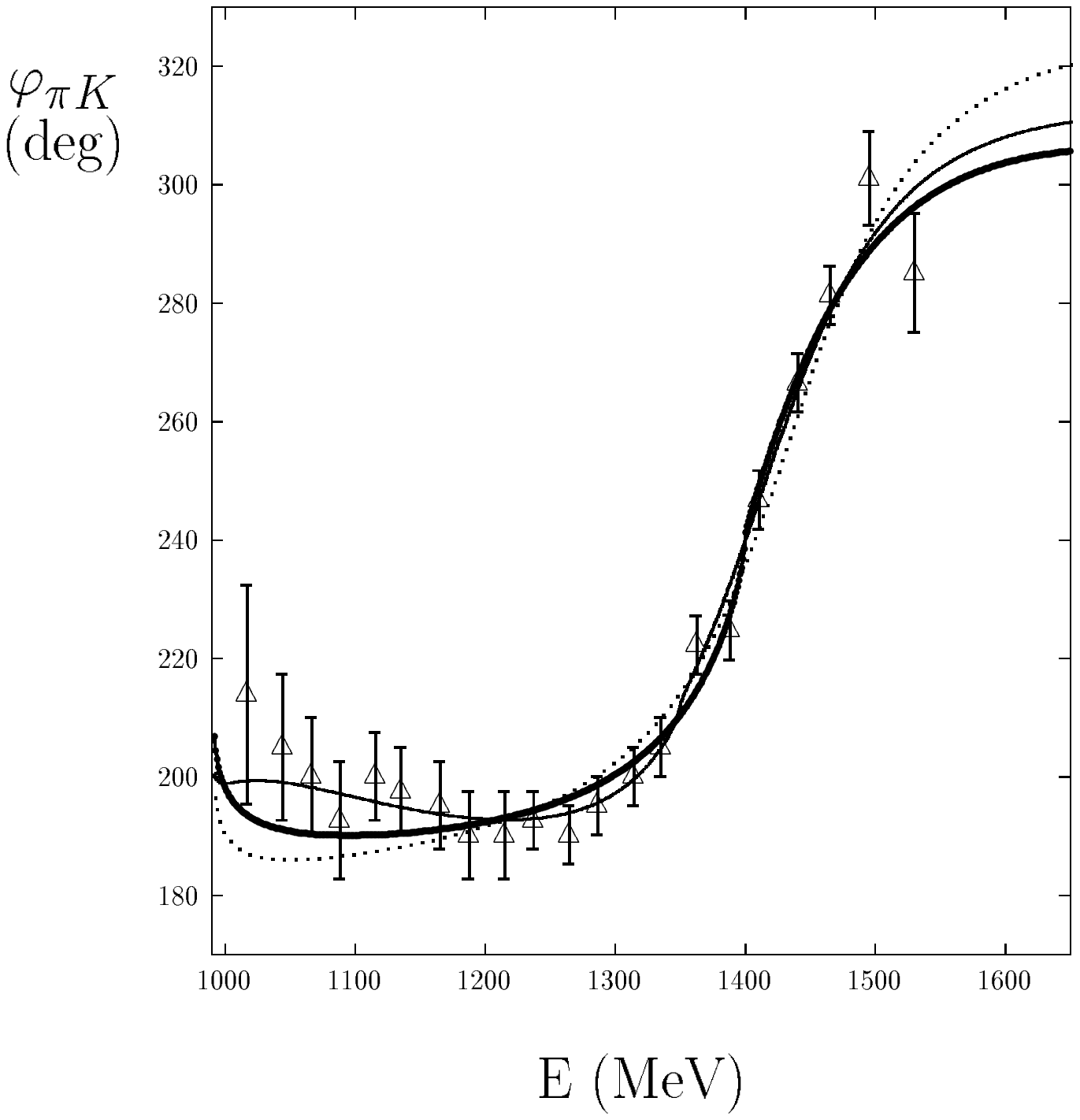}{7cm}{105}{185}{505}{600}{8cm} 
\vspace{0.3cm}
\caption{
Fits to the "down-flat" data of [2],
thick solid line corresponds to fit A, thin solid line to fit B and
dotted line to the 2--channel model fit;
{\bf a)} energy dependence of the $\pi\pi$ phase shifts, 
{\bf b)} energy dependence of inelasticity parameter $\eta_{\pi\pi}$,  
{\bf c)} energy dependence of phase shifts sum
$\varphi_{\pi K} = \delta_{\pi\pi} + \delta_{K\overline{K}}$;
experimental data are from [11].
} 
  \end{figure}
  


\begin{table}[h]
\centering
\caption{Separable interaction parameters for 2-- and 3--channel model fits
to the "down-flat" data from [2].
Values of $\beta$ and $m_3$ are given in GeV.
We use dimensionless coupling
constants defined as 
$\Lambda_{\alpha\gamma,\,j} =
\lambda_{\alpha\gamma,\,j}/2\left(\beta_{\alpha,\,j} 
\beta_{\gamma,\,j}\right)^{3/2}$.}
\begin{tabular}{lccc}
\multicolumn{1}{c}{model} &
\multicolumn{1}{c}{2-channel}       &   
\multicolumn{2}{c}{3--channel} \\
\tableline
\multicolumn{1}{c}{fit}        &
\multicolumn{1}{c}{}        &   
\multicolumn{1}{c}{A}       &
\multicolumn{1}{c}{B}       \\   
\tableline
$\Lambda_{11,1}$ & $-.14258\times 10^{-3}$ & $-.29975\times 10^{-2}$ & $-.52138\times 10^{-2}$  \\
$\Lambda_{11,2}$ & $-.18895$         & $-.10844$         & $-.10552$          \\
$\Lambda_{22}$   & $-.49106$         & $-.39304$         & $3.1637$           \\
$\Lambda_{33}$   & 0                 & $-.17447\times 10^{-2}$ & $-.45719\times 10^{-1}$  \\
$\Lambda_{12,1}$ & $.27736\times 10^{-5}$  & $.12039\times 10^{-2}$  & $.10685\times 10^{-1}$   \\
$\Lambda_{12,2}$ & $.43637\times 10^{-1}$  & $-.11333$         & $-.80626$          \\
$\Lambda_{13,1}$ & 0                 & $-.25841\times 10^{-2}$ & $.14544\times 10^{-4}$   \\
$\Lambda_{13,2}$ & 0                 & $.39924$          & $.21878\times 10^{-2}$   \\
$\Lambda_{23}$   & 0                 & $-.57955$         & $-.17515\times 10^{-1}$  \\
$\beta_{1,1}$    & $3.0233\times 10^3$     & $1.426\times 10^{2}$    & $.81615\times 10^2$      \\
$\beta_{1,2}$    & $1.0922$          & $.92335$          & $.85776$           \\
$\beta_2$        & 2.2941            & $1.4959$          & $.47403$           \\
$\beta_3$        & --------          & $1.3676$          & $.45357\times 10^2$      \\
$m_3$            & --------          & .70               & .67510             \\
\end{tabular}
\label{parameters}
\end{table}    

   
\begin{table}[h]
\centering
\caption{Masses and widths of resonances found for the 2-- and 
3--channel fits to the "down--flat" data}

\begin{tabular}{lccccc}
\multicolumn{1}{c}{pole}         &
\multicolumn{1}{c}{sheet}        &
\multicolumn{2}{c}{A}            &
\multicolumn{2}{c}{B}            \\ 
\cline{3-6}
\multicolumn{1}{c}{}                &
\multicolumn{1}{c}{}                &
\multicolumn{1}{c}{$M$ (MeV)}       &
\multicolumn{1}{c}{$\Gamma$ (MeV)}  & 
\multicolumn{1}{c}{$M$ (MeV)}       &
\multicolumn{1}{c}{$\Gamma$ (MeV)}  \\ 
\tableline
\multicolumn{1}{c}{$f_0(500)$ ($\sigma$)}    & $-++$ & 518.1  & 521.4 & 511.8  & 532.6 \\  
\multicolumn{1}{c}{$f_0(980)$}               & $-++$ & 989.0  & 62.0  & 992.4  & 68.2  \\  
\multicolumn{1}{c}{$f_0(1400)$:}             & $---$ & 1405.1 & 147.8 & 1411.5 & 169.3 \\  
\multicolumn{1}{c}{}                         & $--+$ & 1456.4 & 93.3  & 1402.7 & 108.2 \\  
\end{tabular}
\label{resonances}
\end{table}





\end{document}